\documentclass[12pt]{article}
\pdfoutput=1

\usepackage{amsmath}
\usepackage{amsfonts}
\usepackage{amscd}
\usepackage{amsthm}
\usepackage{setspace}
\usepackage{amssymb}

\usepackage{graphicx}
\usepackage{authblk}
\usepackage{caption}
\usepackage{ytableau}
\usepackage{mathtools}

\begin{document}

\begin{titlepage}

\begin{flushright}

\end{flushright}

\vskip 1cm

\begin{center}

{\bf \Large Black hole graviton and quantum gravity}

\vskip 1.2cm

Yusuke Kimura$^1$ 
\vskip 0.4cm
{\it $^1$Analytical quantum complexity RIKEN Hakubi Research Team,\\ RIKEN Center for Quantum Computing (RQC), 2-1 Hirosawa, Wako, Saitama 351-0198, Japan}
\vskip 0.4cm

\vskip 1.5cm
\abstract{Drawing from a thought experiment that we conduct, we propose that a virtual graviton gives rise to a black hole geometry when its momentum surpasses a certain threshold value on the Planck scale. This hypothesis implies that the propagator of a virtual graviton, that possesses momentum surpassing this threshold, vanishes. Consequently, a Feynman diagram containing this type of graviton propagator does not add to the overall amplitude. This mechanism suggests the feasibility of formulating an ultraviolet-finite four-dimensional quantum gravitational theory. The elementary particles including the gravitons are treated as point particles in this formulation. 
}

\end{center}
\end{titlepage}

\tableofcontents
\section{Introduction}
Quantizing Einstein's general relativity (GR) by incorporating a graviton, which is a boson that mediates the gravitational force, into the framework of quantum field theory is a fundamental approach to formulating the quantum gravitational theory. However, a serious problem is known to arise when attempting to construct this theory. Currently, four-dimensional quantum gravity is considered nonrenormalizable when describing the elementary particles as points particles, resulting in computations that yield meaningless infinite values at the quantum field theory level, given that gravitons are treated as point particles. The three fundamental forces, excluding gravity, all have the problem of infinity resolved through the renormalization. However, when it comes to gravity attempting to quantize a graviton, the ultraviolet (UV) divergence arises, and the infinity cannot be removed using the renormalization method. 

Essentially, the problem of infinity with quantum gravity arises when the two endpoints of a graviton propagator approach to coincide. This occurs when a graviton is described as a point particle. Does this mean that we should abandon the description of a graviton as a point particle when constructing a consistent theory of quantum gravity? Our answer to this question is no. We demonstrate in this study that it is possible to construct a consistent theory of quantum gravity without discarding the notion of graviton as a point particle. 

Generally, a graviton is described as an elementary bosonic, massless and spin 2 particle. Undoubtedly, gravitons possess all these properties. However, do these properties exhaustively describe the characteristic features of gravitons? We believe that one important property of gravitons is overlooked in this description. The omission of this fundamental aspect of gravitons hinders one from constructing a UV-finite quantum gravitational theory where a graviton is described as a point particle. 

We consider the unnoticed property of graviton, a new aspect of graviton in a sense, as its essential and characterizing property. We propose in this note that using this new aspect of graviton will provide a way to resolve the problem of UV divergence in quantum gravity. 

As part of our strategy, we will demonstrate that a graviton possesses an essential and intrinsic property beyond just being a massless, spin 2 bosonic elementary particle, by performing a thought experiment. The nature of a graviton cannot be precisely captured without the property that we specify in this work. To identify the unique characteristic property that only gravitons possess, we consider the specific aspect that sets gravity apart from the other three fundamental forces. Gravity changes the spacetime metric (or, to be more precise, the curvature of the spacetime metric manifests as gravity), and under certain conditions, a black hole forms owing to the effect of gravity. This is a characteristic property of gravity, which the other three fundamental forces lack. Particularly, the formation of a black hole is specific to gravity, and this phenomenon does not occur with the other three forces. 

This observation leads us to an idea that a graviton can also have the same property. Namely, it is likely that, under certain conditions, a graviton is able to change the spacetime metric, resulting in the formation of the horizon of a black hole geometry that surrounds the graviton itself. Our thought experiment suggests that gravitons indeed have this property as we will show shortly. In Section \ref{subsec2.1}, we conducted a thought experiment that provides evidence for gravitons possessing this property. The implications of the thought experiment that we conducted in Section \ref{subsec2.1} are discussed in Section \ref{subsec2.2}. 

The thought experiment, computation, and physical insight that we utilize in this study suggest that the phenomenon of a graviton generating a black hole geometry around it occurs only at the Planck scale.

Virtual gravitons can possess the feature to form a black hole geometry, and this does not seem to apply to other types of {\it actual} elementary particles. We will discuss the new aspect of a graviton to generate a black hole geometry around itself, and based on this property, we will propose a quantum gravitational theory. 

In Section \ref{subsec3.1}, we hypothesize a spacetime metric equation for the black hole geometry generated by a virtual graviton. We discuss the physical consequences and implications in Section \ref{subsec3.2}. 

Gravity is mediated through an exchange process of virtual gravitons, emitted from one particle and absorbed by another particle. The mechanism that we hypothesize in Section \ref{sec3} implies that the graviton propagator vanishes for any momentum higher than (a constant of $\mathcal{O}(1)$ times) the Planck momentum. Physical consequences of this will be discussed in Section \ref{subsec3.2}. Owing to this property of gravitons, any length shorter than the Planck scale does not have a meaning for the gravitational interactions. As a result of this mechanism, quantum field theory incorporating a graviton does not have a UV divergence, resulting in a consistent quantum gravitational theory. 

In this study, we propose a hypothesis that when a virtual graviton possesses a Planck scale momentum larger than a specific value, the spacetime metric around the graviton becomes that of the Kerr black hole \cite{Kerr1963}. The details of this will be described in Section \ref{subsec3.1}. We refer to the graviton generating the Kerr black hole geometry as the ``black hole graviton.'' In Section \ref{sec3}, we will clarify that a virtual graviton with a large momentum can naturally give rise to a black hole geometry. The quantum gravitational aspect becomes important for the black hole graviton. 

Our hypothesis, proposed in this work, implies that once a virtual graviton exceeds the threshold Planck scale momentum, it generates a black hole geometry forming a horizon, evaporating instantly owing to the Hawking radiation \cite{Hawking1975}. These consequences suggest that constructing a UV-finite quantum gravitational theory that treats gravitons as point particles is possible. 

We also hypothesize that below the threshold momentum, the spacetime metric around a virtual graviton is well approximated by the flat Minkowski metric, and the metric does not transition to the Kerr geometry below the threshold value.  

Discussions to treat particles as spacetime singularities can be found in \cite{Einstein1938, Einstein1940}. 

In Section \ref{subsec3.1}, we estimate the lifetime of a black-hole graviton by utilizing the energy-time uncertainty principle. An estimation of the lifetime of a virtual particle using the energy-time uncertainty relation can be found in \cite{Yukawa1935}. We state our concluding remarks in Section \ref{sec4}.

We utilize the covariant perturbation method to quantize the gravitational theory. Perturbative analyses of quantum gravity can be found, e.g., in \cite{Feynman1963, DeWitt19672, DeWitt19673, Faddeev1967, Weinberg1980}.

\section{Thought experiment of black hole geometry}
\label{sec2}
\subsection{Thought experiment}
\label{subsec2.1}
One of the characteristic properties of a black hole geometry is that the geometry produces a ``horizon'' enclosing a bounded region in the spacetime. Once a light is inside the horizon, it cannot escape to the outside. Is it possible to represent this property of a black hole using the equivalence principle? 

To represent the black hole geometry using the equivalence principle, we conduct the following thought experiment. 

Let's consider a small room with a height of $h$, where no gravity is acting. The room is accelerating upward with a constant acceleration, $g$. Two observers are situated in this room - observer A at the bottom, $z=0$, and observer B at the top, $z=h$. At time $t=0$, observer A emits a signal with the speed of light toward the observer B. It takes at least $\frac{h}{c}$ for the signal to reach observer B, regardless of the value of the acceleration $g$ (where we assume that the acceleration is nonnegative, $g\ge 0$). When the acceleration is nonzero, the signal needs time strictly longer than $\frac{h}{c}$. 

When time $\frac{h}{c}$ have elapsed, the emitted signal moves upward by $h$, while observer A moves upward by $\frac{1}{2}g(\frac{h}{c})^2$. Therefore, if the acceleration $g$ satisfies the following relation:
\begin{equation}
\label{ineq_1_sec2}
\frac{1}{2}\,g\left(\frac{h}{c}\right)^2\ge h,
\end{equation}
the signal returns to the observer A who emitted the signal, meaning that the emitted signal never reaches observer B. 

Because the speed of the room reaches the speed of light after a certain time, this thought experiment makes sense only for a limited time interval. However, the thought experiment represents a situation where light cannot escape from the strong gravitational field. 

An important aspect of the thought experiment is that this represents a situation where noting can escape from inside the horizon of a black hole geometry to the outside. 

Indeed, one can derive the characteristic coefficient of the metric for the Schwarzschild black hole \cite{Schwarzschild1916} (up to a constant) from this thought experiment. We would like to demonstrate this point. 

The inequality \eqref{ineq_1_sec2} that we just deduced in the thought experiment can be rewritten as follows:
\begin{equation}
\label{ineq_2_sec2}
\frac{gh}{2c^2}\ge 1.
\end{equation}
In this form, the factor, $gh$, in the inequality \eqref{ineq_2_sec2} can be viewed using the equivalence principle as the difference in the gravitational potential, $\Delta \Phi$, at the bottom, $z=0$, and the top, $z=h$, of the room. Therefore, the inequality \eqref{ineq_2_sec2} can also be rewritten as the following:
\begin{equation}
\label{ineq_3_sec2}
\frac{\Delta\Phi}{2c^2}-1\ge 0.
\end{equation}
The gravitational potential $\Phi$ can be expressed as the following:
\begin{equation}
\Phi(r)=-\frac{GM}{r}.
\end{equation}
One can consider a situation where the distance, $r$, of the source of the gravitational field to the room is comparable to the Planck scale. In this situation, because the size of the room must be smaller than this distance, one has the following relation: $h\le r$. However, any length far below the Planck length is considered not to have meaning. If $r$ is comparable to the Planck length $l_P$, it means that $h$ is also comparable to the Planck length, i.e. $h\sim r$. 

These considerations imply that in this situation, the difference in the gravitational potential, $\Delta \Phi$, takes the following form:
\begin{equation}
\Delta\Phi=-\left(\frac{GM}{r+h}-\frac{GM}{r}\right)\sim \frac{GM}{r}-\frac{GM}{r+r}=\frac{GM}{2r}.
\end{equation}
By plugging this expression in, the inequality \eqref{ineq_3_sec2} becomes as follows:
\begin{equation}
\frac{GM}{4c^2r}-1\ge 0.
\end{equation}
This reproduces the coefficient of the $(cdt)^2$ component of the metric for the Schwarzschild geometry up to a constant \footnote{The constant coefficient of $\frac{GM}{c^2r}$ should be 2, not $\frac{1}{4}$, for the precise Schwarzschild geometry expression. However, since the difference in the constants can be absorbed into $G$, this difference is inconsequential.}. 

\subsection{Implications of thought experiment}
\label{subsec2.2}
We would like to discuss the implication of the outcomes of this thought experiment. The thought experiment represents a black hole geometry, and a characteristic aspect of a black hole geometry is visible, particularly when the thought experiment is applied to the Planck scale-sized system. This means that this thought experiment does not describe a celestial black hole, but rather describes a black hole geometry on the Planck scale. These series of considerations strongly suggest that a black hole formation indeed occurs in the regime of elementary particles. Which elementary particle has the property to change the spacetime metric to generate a black hole geometry? Because forming a black hole is an intrinsic property of gravity, it is natural to expect that a graviton, which mediates the gravitational force, has the property. 

In contrast, applying this thought experiment to {\it actual} elementary particles other than gravitons seems physically implausible for the following reason: without the presence of gravitons, the mass of an elementary particle is just a number. The mass of the particle has physical meaning and consequence only when it couples with graviton. At the elementary particle level, a force is mediated through the exchange of force-carrying bosons. However, in the situation described in the thought experiment we just considered, any signal emitted from an elementary particle with a black hole geometry cannot escape to the outside of the horizon. This argument applies to the force-carrying boson; therefore, the exchange of virtual bosons does not occur for the elementary particle that generates a black hole geometry. Elementary particles, excluding gravitons, do not contribute to gravity without interacting with gravitons. Thus, it is physically implausible to expect that actual elementary particles, excluding gravitons, form a black hole geometry by themselves.   

Given this, we focus on the possibility that (virtual) gravitons possess the property to generate a black hole geometry on their own. The arguments that we presented previously do not necessarily rule out the possibility that other types of virtual elementary particles also have the ability to generate a black hole geometry.

Here, we propose that the virtual graviton generates a black hole geometry. While an actual on-shell graviton is massless, a virtual graviton possessing timelike momentum can be thought of as a ``massive'' particle. When a virtual graviton has momentum $p$, this can be interpreted as it having mass $\frac{|p|}{c}$. We introduced the notation $|p|$, which is defined by the equation:
\begin{equation}
\label{def_p}
p^2=p_{\mu}p^{\mu}=-|p|^2.
\end{equation}
In this note, for simplicity we focus on the situation where momentum of a virtual graviton is timelike or lightlike.  

Since a graviton has spin 2, if a virtual graviton gives rise to a black hole geometry, it must be described by the Kerr geometry; the mass parameter, $M$, in the Kerr geometry is naturally associated with ``mass'' $\frac{|p|}{c}$ of the virtual graviton. It is reasonable to expect that a virtual graviton generates a black hole geometry when its momentum exceeds a Planck-scale threshold. As we will discuss in the next section, this expectation is physically plausible. 

The reasoning that we gave previously has an important implication. Based on the thought experiment we conducted, we discussed that virtual gravitons of high-momentum modes \footnote{In this study, we utilize the following convention for the Minkowski metric:
\begin{equation}
\eta_{\mu\nu} = \begin{pmatrix}
-1 & 0 & 0 & 0 \\
0 & 1 & 0 & 0 \\
0 & 0 & 1 & 0 \\
0 & 0 & 0 & 1 \\
\end{pmatrix}.
\end{equation}
For a four-momentum $p_{\mu}$, we refer to its state as ``high'' and ``large'' when the value of $-p^2=-p_\mu p^\mu$ is large.} will generate a black hole geometry. Consequently, the wave functions of the virtual gravitons with high momenta are concealed by the horizons of the black hole geometries. This implies that the graviton propagator becomes zero at large momenta. Since a virtual particle in high-momentum mode represents a quantum fluctuation at a short distance, this can be restated as: the influence of the graviton propagator of a virtual graviton vanishes at sufficiently short distances. This consideration leads us to propose that the gravitational coupling, $g_{\rm Grav.}$, vanishes at large momenta: 
\begin{equation}
g_{\rm Grav.}(p)=0 \hspace{.2in} 
\end{equation}
when $-p^2$ exceeds a certain threshold value. This will be discussed in detail in the next section. We will find that this happens at the Planck scale.

As we will discuss shortly, the postulate that we propose in this work implies that the graviton propagator vanishes for virtual gravitons with momenta higher than the Planck scale.  

In the previous thought experiment and discussions, the following three points are particularly important:
\begin{itemize}
\item[] i) In a thought experiment, (using the equivalence principle, on the side where the gravitational force is acting) the experimental room should be chosen to be sufficiently small compared to the size of the region on which the gravitational field acts. However, when the region on which the gravitational field acts is chosen to be the Planck scale size, since a distance shorter than the Planck length scale does not have any meaning in principle, the size of the room in the experiment must be on the order of the Planck length. \\
\item[] ii) A virtual graviton can be seen as a ``massive'' particle, with its momentum being considered its ``mass'' (times $c$). This ``mass'' of the virtual graviton corresponds to the mass parameter of the black geometry that it generates. \\
\item[] iii) The graviton propagator becomes zero for gravitons of sufficiently high-momentum modes. As a result, the gravitational coupling, $g_{\rm Grav.}$, vanishes at large momenta.
\end{itemize}

\section{Black hole graviton}
\label{sec3}
\subsection{Our proposed hypotheses}
\label{subsec3.1}
In the previous section, we conducted a thought experiment that suggested that the spacetime metric around an elementary particle can become that of a black hole geometry. By physical reasoning, we also deduced that if this occurs, the elementary particle must be the graviton. 

When a graviton is treated as a point particle, four-dimensional quantum gravitational theory is considered nonrenormalizable. We argue that this results from overlooking an important property of gravitons. Owing to this oversight, the theory does not correctly include this aspect of the gravitons, leading to meaningless infinite results in computations. Here, we demonstrate that if a graviton gives rise to a black hole geometry, it results in a UV-finite four-dimensional quantum gravitational theory. To be more precise, we will demonstrate that when a graviton is treated as a point particle, incorporating an unnoticed aspect of graviton's property into the quantum gravitational theory makes the four-dimensional quantum gravity UV finite. 

Our proposal that gravitons generate a black hole geometry under a certain condition yields a UV-finite quantum gravitational theory, providing strong support for our proposal. Given that generating a black hole geometry under a certain condition is an intrinsic nature of gravity, it is natural to consider that gravitons possess the same property. Constructing a consistent quantum gravitational theory is possible when incorporating this property of a graviton. 

In quantum field theory, an interaction that transmits a force occurs through the exchange of virtual particles between particles. However, when describing elementary particles as point particles, a virtual particle travels between two spacetime points in the Feynman diagram. As these two endpoints of the virtual particle propagator come closer together, approaching coincidence, the problem of divergence inevitably arises. This is the origin of divergences in quantum field theory computations, including UV divergence that arises when quantizing the gravitational theory.
 
Therefore, if there is a mechanism to prevent virtual gravitons from travelling with momenta higher than a certain specific value, the problem of divergence does not arise in quantizing the gravitation.

We propose a mechanism in which graviton propagators possessing high momenta that exceed a certain threshold will vanish. Graviton is an electrically neutral, spin 2 boson. Therefore, if a virtual graviton generates a black hole geometry, it must be described as a Kerr geometry as we stated previously. 

Based on the thought experiment and the preceding discussions given here and in Section \ref{sec2}, we propose the following hypothesis: \\
Virtual graviton possessing momentum $p$, with $-p^2$ surpassing a threshold value, results in a black hole geometry described by the following metric:
\begin{align}
\label{graviton_metric}
ds^2 & = -\left(1-\frac{\frac{2G|p|}{c^3}r}{r^2+a^2\, {\rm cos}^2\theta}\right) (cdt)^2 - \frac{\frac{4G|p|}{c^3}ra\,{\rm sin}^2\theta}{r^2+a^2\, {\rm cos}^2\theta } (cdt)d\phi + \frac{r^2+a^2\, {\rm cos}^2\theta}{r^2+a^2-\frac{2G|p|}{c^3}r}dr^2 \\\nonumber
& +\left(r^2+a^2\, {\rm cos}^2\theta \right)d\theta^2 + \left(r^2+a^2+\frac{\frac{2G|p|}{c^3}ra^2}{r^2+a^2\, {\rm cos}^2\theta}{\rm sin}^2 \theta\right) {\rm sin}^2 \theta d\phi^2.
\end{align}
The coordinates $(t,r,\theta,\phi)$ represent Boyer--Lindquist coordinates \cite{Boyer1966}. We have introduced $|p|$ in Section \ref{subsec2.2} defined by the equation \eqref{def_p}, and $a$ represents the following quantity:
\begin{equation}
a=\frac{2\hbar}{|p|}.
\end{equation}
$\frac{|p|}{c}$ can be considered as the ``mass'' of the virtual graviton. We will provide a description of the threshold value mentioned in the statement shortly. 

The thought experiment conducted in Section \ref{sec2} suggested that a virtual graviton generates a black hole geometry when its momentum exceeds a certain threshold value. Does the metric \eqref{graviton_metric} physically include this threshold value? 

One can confirm that the metric possesses this property by analyzing the black hole geometry \eqref{graviton_metric}. As for the Kerr geometry, a condition is imposed on the angular momentum. When the condition is satisfied, a horizon forms that makes the metric a black hole geometry. The condition is given by the following inequality:
\begin{equation}
\label{ineq_Kerr}
J\le \frac{M^2G}{c}.
\end{equation}
If this condition is not met, the issue of naked singularity arises. 

In the situation where the Kerr geometry is applied to a virtual graviton, the graviton spin yields the angular momentum, $J=2\hbar$, and $M$ is replaced by $\frac{|p|}{c}$; therefore, the condition \eqref{ineq_Kerr} becomes as follows:
\begin{equation}
\label{ineq_Kerr_2}
2\hbar\le \frac{|p|^2G}{c^3}.
\end{equation}
This is equivalent to the following condition:
\begin{equation}
\sqrt{\frac{2\hbar c^3}{G}}\le |p|.
\end{equation}
Using the Planck length $l_P$, this can be rewritten as follows:
\begin{equation}
\frac{\sqrt{2}\hbar}{l_P}\le |p|.
\end{equation}
Furthermore, using the Planck momentum $P_{\rm Planck}=\frac{\hbar}{l_P}$, the condition \eqref{ineq_Kerr_2} can also be rephrased as follows:
\begin{equation}
\sqrt{2}P_{\rm Planck}\le |p|.
\end{equation}
If momentum of the virtual graviton does not exceed the threshold value, the problem of naked singularity arises for the metric \eqref{graviton_metric}.

Since the appearance of naked singularity is undesirable based on the cosmic censorship conjecture \cite{Penrose1969}, we also propose that a virtual graviton only gives rise to the spacetime metric \eqref{graviton_metric} when its momentum surpasses the threshold value, $\frac{\sqrt{2}\hbar}{l_P}=\sqrt{2}P_{\rm Planck}$. Under this assumption, the issue of naked singularity does not arise. 

\vspace{.2in}
{\it Remark}: \\
A virtual graviton is a virtual particle; therefore, this will not be detectable. This means that the process of a virtual graviton giving rise to a black hole geometry and then it evaporating completely to disappear should also be undetectable. The energy-time uncertainty relation can explain this phenomenon.  

Utilizing the Komar integral \cite{Komar1958, Cohen1984} \footnote{This quantity coincides with the Arnowitt--Deser--Misner (ADM) energy \cite{Arnowitt1959, Arnowitt1962} for a Kerr geometry.}, the energy of the virtual graviton with $-p^2=m^2c^2$ is interpreted as $mc^2$. Thus, the energy $\Delta E$ of a virtual graviton with its momentum at the threshold value is estimated as follows:
\begin{equation}
\Delta E \sim \sqrt{2}P_{\rm Planck}c=\frac{\sqrt{2}\hbar c}{l_P}.
\end{equation}
Therefore, if the virtual graviton completely evaporates within a time frame
\begin{equation}
\label{time_bound}
\Delta t \le \frac{\hbar}{2\Delta E}=\frac{l_P}{2\sqrt{2}c}=\frac{t_P}{2\sqrt{2}},
\end{equation}
this virtual process becomes undetectable, based on the energy-time uncertainty relation. $t_P$ in \eqref{time_bound} represents the Planck time. 

The ``mass'' of the virtual graviton generating the black hole geometry is on the Planck scale. This means that the lifetime estimation formula, $\tau\sim M^3$, \cite{Hawking1975, Page1976, Page19762} for black hole evaporation owing to the Hawking radiation, which is derived assuming that the mass of a black hole is well above the Planck scale, does not apply to this phenomenon.

Based on these considerations, we expect that the virtual gravitons with momenta surpassing the threshold value will evaporate within a timeframe relative to the Planck scale. 

\subsection{Implications of the hypotheses}
\label{subsec3.2}
We proposed that a virtual graviton with momentum exceeding the threshold value, $\frac{\sqrt{2}\hbar}{l_P}$, generates the black hole geometry \eqref{graviton_metric}. We will now demonstrate that this hypothesis leads to a UV-finite quantum gravitational theory. 

The exchange of virtual gravitons and the graviton propagator mediate the gravitational interaction, and these are described by the Feynman diagram. The polarization tensor resulting from a metric perturbation describes the wave function of a graviton. 

For a virtual graviton in a state with its momentum smaller than the threshold value $\frac{\sqrt{2}\hbar}{l_P}$, the virtual graviton contributes to the amplitude through the graviton propagator, transferring momentum.

However, when the momentum of a virtual graviton surpasses the threshold value, the situation changes. Once a virtual graviton with momentum exceeding the value $\frac{\sqrt{2}\hbar}{l_P}$ is emitted from one endpoint of the propagator, the spacetime metric around the virtual graviton becomes that of \eqref{graviton_metric}. As a result, a horizon forms, enclosing the virtual graviton. The resulting horizon $r_+$ conceals the virtual graviton and prevents its polarization tensor $\gamma_{\mu\nu}$ from transmitting outside of the horizon. Consequently, the polarization tensor $\gamma_{\mu\nu}$ of the virtual graviton outside the horizon $r_+$ is zero. Therefore, the virtual graviton with momentum surpassing the value $\frac{\sqrt{2}\hbar}{l_P}$ does not propagate between the two endpoints of the propagator, which means that the graviton propagator vanishes once its momentum exceeds the threshold value $\frac{\sqrt{2}\hbar}{l_P}$. The virtual graviton is hidden from the other processes of the Feynman diagram that occur outside the horizon. Since this diagram involves the graviton propagator which has a value of zero, the contribution of this Feynman diagram is also zero. This means that the Feynman diagram involving a virtual graviton with momentum surpassing the threshold value $\frac{\sqrt{2}\hbar}{l_P}$ does not contribute to the whole amplitude. 

In essence, this is equivalent to stating that for lengths shorter than the distance $\frac{l_P}{\sqrt{2}}$, virtual gravitons do not contribute through a propagator. 

Since the formation of the horizon occurs on the Planck scale for a virtual graviton, we expect that the resulting black hole evaporates completely owing to Hawking radiation on a timescale comparable to the Planck time $t_P$. 

From these arguments, we find that in a high-momentum state exceeding the threshold value,
the gravitational coupling $g_{\rm Grav.}(p)$ vanishes, i.e.
\begin{equation}
g_{\rm Grav.}(p)=0
\end{equation}
for
\begin{equation}
-p^2 \,\textgreater \,\frac{2\hbar^2}{l_P^2}.
\end{equation}

We have deduced that the propagator of a virtual graviton with momentum surpassing the threshold value vanishes, and the gravitational coupling yields a value of zero in such high-momentum states surpassing the threshold. These properties of virtual gravitons make the quantum gravitational theory UV-finite. 

\section{Concluding remarks}
\label{sec4}
In this study, we proposed that a virtual graviton, exceeding a threshold momentum, generates a Kerr black hole geometry. This phenomenon occurs when $-p^2$ surpasses $\frac{2\hbar^2}{l_P^2}$. The resulting black hole is expected to evaporate completely owing to Hawking radiation within a timeframe relative to the Planck time, $t_P$. In these high-momentum states, the polarization tensors of the virtual gravitons are concealed from the external domain by the formed horizons, and the graviton propagator vanishes. 

The momentum threshold beyond which a virtual graviton forms a black hole geometry is on the Planck scale. This observation aligns with the expectation that quantum gravitational effect becomes significant at this scale. 

We predict that a black hole geometry generated by a virtual graviton with momentum exceeding the threshold value will evaporate completely within a timeframe comparable to the Planck time, in accordance with the energy-time uncertainty relation. 

\section*{Acknowledgments}

Y.K. acknowledges Hakubi projects of RIKEN.


\begin{thebibliography}{99}

\bibitem{Kerr1963}
R.~P.~Kerr,
``Gravitational field of a spinning mass as an example of algebraically special metrics,''
{\it Phys. Rev. Lett.} \textbf{11}, 237--238 (1963).

\bibitem{Hawking1975}
S.~W.~Hawking,
``Particle Creation by Black Holes,''
{\it Commun. Math. Phys.} \textbf{43}, 199--220 (1975).

\bibitem{Einstein1938}
A.~Einstein, L.~Infeld and B.~Hoffmann,
``The Gravitational equations and the problem of motion,''
{\it Annals Math.} \textbf{39}, 65--100 (1938).

\bibitem{Einstein1940}
A.~Einstein and L.~Infeld,
``The Gravitational equations and the problem of motion. 2.,''
{\it Annals Math.} \textbf{41}, 455--464 (1940).

\bibitem{Yukawa1935}
H.~Yukawa,
``On the Interaction of Elementary Particles. I,''
{\it Proc. Phys. Math. Soc. Japan} \textbf{17}, 48--57 (1935).

\bibitem{Feynman1963}
R.~P.~Feynman,
``Quantum theory of gravitation,''
{\it Acta Phys. Polon.} \textbf{24}, 697--722 (1963).

\bibitem{DeWitt19672}
B.~S.~DeWitt,
``Quantum Theory of Gravity. 2. The Manifestly Covariant Theory,''
{\it Phys. Rev.} \textbf{162}, 1195--1239 (1967).

\bibitem{DeWitt19673}
B.~S.~DeWitt,
``Quantum Theory of Gravity. 3. Applications of the Covariant Theory,''
{\it Phys. Rev.} \textbf{162}, 1239--1256 (1967).

\bibitem{Faddeev1967}
L.~D.~Faddeev and V.~N.~Popov,
``Feynman Diagrams for the Yang-Mills Field,''
{\it Phys. Lett.} \textbf{B25}, 29--30 (1967).

\bibitem{Weinberg1980}
S.~Weinberg,
``Ultraviolet divergences in quantum theories of gravitation,'' in {\it General Relativity; an Einstein Centenary Survey}, edited by S.~W.~Hawking and W.~Israel, Cambridge University Press (1980).

\bibitem{Schwarzschild1916}
K.~Schwarzschild,
``On the gravitational field of a mass point according to Einstein's theory,''
{\it Sitzungsber. Preuss. Akad. Wiss. Berlin (Math. Phys. )} \textbf{1916}, 189--196 (1916).

\bibitem{Boyer1966}
R.~H.~Boyer and R.~W.~Lindquist,
``Maximal analytic extension of the Kerr metric,''
{\it J. Math. Phys.} \textbf{8}, 265 (1967).

\bibitem{Penrose1969}
R.~Penrose,
``Gravitational collapse: The role of general relativity,''
{\it Riv. Nuovo Cimento} \textbf{1}, 252--276 (1969).

\bibitem{Komar1958}
A.~Komar,
``Covariant conservation laws in general relativity,''
{\it Phys. Rev.} \textbf{113}, 934--936 (1959).

\bibitem{Cohen1984}
J.~M.~Cohen and F.~de Felice,
``The total effective mass of the Kerr-Newman metric,''
{\it J. Math. Phys.} \textbf{25}, 992 (1984).

\bibitem{Arnowitt1959}
R.~L.~Arnowitt, S.~Deser and C.~W.~Misner,
``Dynamical Structure and Definition of Energy in General Relativity,''
{\it Phys. Rev.} \textbf{116}, 1322--1330 (1959).

\bibitem{Arnowitt1962}
R.~L.~Arnowitt, S.~Deser and C.~W.~Misner,
``The Dynamics of general relativity'' in {\it Gravitation: an introduction to current research}, L.~Witten, ed. (Wiley, New York, 1962).

\bibitem{Page1976}
D.~N.~Page,
``Particle Emission Rates from a Black Hole: Massless Particles from an Uncharged, Nonrotating Hole,''
{\it Phys. Rev.} \textbf{D13}, 198--206 (1976).

\bibitem{Page19762}
D.~N.~Page,
``Particle Emission Rates from a Black Hole. 2. Massless Particles from a Rotating Hole,''
{\it Phys. Rev.} \textbf{D14}, 3260--3273 (1976).

\end{thebibliography}
\end{document}